\documentclass[twocolumn,showpacs,preprintnumbers,amsmath,amssymb]{revtex4-1}

\usepackage{graphicx}
\usepackage{dcolumn}
\usepackage{bm}
\usepackage{color}

\begin{document}

\preprint{APS/123-QED}

\title{Avoided ferromagnetic quantum critical point in CeZn}

\author{Hisashi Kotegawa$^{1}$, Toshiaki Uga$^{1}$, Hideki Tou$^1$, Eiichi Matsuoka$^1$, Hitoshi Sugawara$^1$}

\affiliation{
$^{1}$Department of Physics, Kobe University, Kobe 658-8530, Japan \\
}

\date{\today}

\begin{abstract}

Cubic CeZn shows a structural phase transition under pressure, and it modifies the ground state from an antiferromagnetic (AFM) state to a ferromagnetic (FM) state. To investigate how the FM state terminates at a quantum phase transition, we measured the electrical resistivity under pressure for a single crystal CeZn. The transition temperature into the FM state decreases monotonously with increasing pressure, accompanied by the pronounced Kondo effect, but a drastic change in the field response occurs before the ordered phase terminates. This result suggests that the FM quantum critical point is avoided by the appearance of an AFM-like state.

\end{abstract}

\maketitle

In condensed matter physics, a combination of elements and crystal structure brings diversity in the electronic state of materials.
Therefore, universality present in various materials is important.
The ''absence of ferromagnetic (FM) quantum critical point (QCP)'' is an interesting hypothesis to represent such universality.
A QCP, which is a continuous phase transition at zero temperature, has been seen in many systems.
However, in metallic ferromagnets, a QCP at zero field is hardly achieved \cite{Brando,Pfleiderer_MnSi,Uhlarz,Kabeya,Valentin,Kotegawa1,Aoki,Kimura,Shimizu,Araki,Kobayashi,Wilhelm,Sullow,Sidorov,Moroni,Niklowitz,Lausberg,Kotegawa2,Lengyel,Valentin_LaCrGe3,Kaluarachchi,Kaluarachchi2}.
A notable way to prevent an FM QCP is the appearance of a tricritical point (TCP), at which the second-order FM transition changes to a first-order transition \cite{Pfleiderer_MnSi,Uhlarz,Kabeya,Valentin,Kotegawa1,Aoki,Kimura,Shimizu,Araki}.
The theoretical work has specified its universality; that is, the TCP appears irrespective of the electronic structure \cite{Brando,Belitz}.
Another way to prevent an FM QCP is by switching of the ordered state.
In many systems, the FM transitions are preempted by the transition into a different phase such as an antiferromagnetic (AFM) phase, before the QCP is reached \cite{Brando,Kobayashi,Wilhelm,Sullow,Sidorov,Moroni,Niklowitz,Lausberg,Kotegawa2,Lengyel,Valentin_LaCrGe3,Kaluarachchi,Kaluarachchi2}.
The switching of the ordered state can be caused by a change in the magnetic interaction or a formation of a modulated magnetic phase induced by quantum fluctuation \cite{Conduit,Efremov,Karahasanovic,Abdul}.

However, it was found that CeRh$_6$Ge$_4$, which crystalizes in the hexagonal structure with the space group $P$\=6$m$2, is a possible counterexample \cite{Matsuoka,Kotegawa_CeRh6Ge4,Shen}.
It has been suggested that the Curie temperature of 2.5 K reaches a QCP under pressure at zero field. 
Although microscopic measurements are highly required to confirm the QCP, noncentrosymmetry might be a key ingredient to induce an FM QCP \cite{Kirkpatrick}.
These studies show the importance of surveying FM materials to clarify what determines the behavior in the boundary of the FM QCP in the actual material.

Meanwhile, the difficulty of such an investigation arises from the fact that metallic ferromagnets near the quantum phase transition are few.
To investigate the intrinsic behavior of the FM quantum phase transition, high-quality crystals are essential \cite{Sang}.
In this study, we focus on CeZn that satisfies these conditions.
The CeZn crystallizes in a cubic structure with the space group $Pm\bar{3}m$ at ambient pressure \cite{Chao}.
It shows an AFM transition at $T_N=30$ K \cite{Schmitt}, whereas an FM transition appears under pressure, accompanied by a structural phase transition \cite{Kadomatsu,Shigeoka}.
It has been reported that CeZn transforms into a rhombohedral structure above $\sim2.5$ GPa at 300 K and $\sim1.0$ GPa at low temperatures \cite{Shigeoka,Uwatoko}, but a recent measurement revealed that it is a tetragonal structure with space inversion symmetry ($P4/nmm$) \cite{XShen}.
The pressure-temperature phase diagram up to $\sim3$ GPa has been reported in earlier studies \cite{Kadomatsu,Shigeoka}, in which the FM phase survives even at the highest pressure, and a recent study has confirmed that the magnetically ordered state disappears above $\sim3$ GPa \cite{XShen}.

\begin{figure}[htb]
\centering
\includegraphics[width=0.8\linewidth]{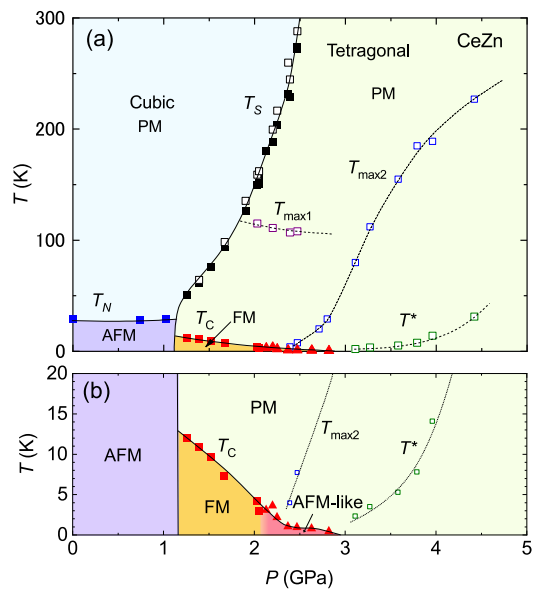}
\caption[]{(color online) Pressure-temperature phase diagram of CeZn, which is obtained in this study. (a): $0<T<300$ K, (b): $0<T<20$ K. The AFM phase at ambient pressure changes into the FM phase through the cubic to tetragonal structural phase transition. The magnetically ordered state terminates at $P_c \sim3$ GPa. The metamagnetic-like transitions appear in the pressure region, where the transition temperatures are indicated by the closed triangles. They suggest that the FM state changes into an AFM-like state just before a QCP is reached. }
\end{figure}

In this study, we investigated the quantum phase transition of CeZn using resistivity measurements under pressure up to $\sim4.5$ GPa and under magnetic fields.
To give an outlook of this study, the obtained phase diagram is shown in Fig.~1.
The first-order structural phase transition at $T_S$ and the modification from the AFM state to the FM state are consistent with those of previous studies \cite{Kadomatsu,Shigeoka,Uwatoko}.
In contrast to the previous study \cite{Kadomatsu}, the FM transition temperature $T_C$ in the tetragonal phase decreases gradually at increasing pressure.
The quantum phase transition into the paramagnetic (PM) phase likely occurs at $P_c \sim 3$ GPa.
These results are consistent with those of the recent report \cite{XShen}.
We found that the metamagnetic-like transition suddenly emerges above $\sim2.1$ GPa, indicating that the ordered state is no longer the FM state.
The experimental observation suggests that the FM state transforms into another AFM-like state just before a QCP is reached in CeZn.

A single crystal of CeZn was made using a similar procedure to that in Ref.~\cite{Shigeoka}.
The residual resistivity ratio at ambient pressure is approximately 60, ensuring a high-quality crystal.
High pressure was applied using an indenter-type pressure cell \cite{indenter}, and Daphne7474 was used as a pressure-transmitting medium \cite{Murata}.
The low temperatures down to 0.4 K were achieved using a $^3$He cryostat.
For electrical resistivity measurement, the current flowed along the cubic [110] direction, and in most cases the magnetic field was applied along the cubic [$\bar{1}$10] direction ($j \perp H$).
However, the sample might be in a multidomain in the tetragonal phase.
The cubic [110] and [$\bar{1}$10] directions correspond to any of the tetragonal [100], [112], and their equivalent directions \cite{XShen}.

\begin{figure}[htb]
\centering
\includegraphics[width=1\linewidth]{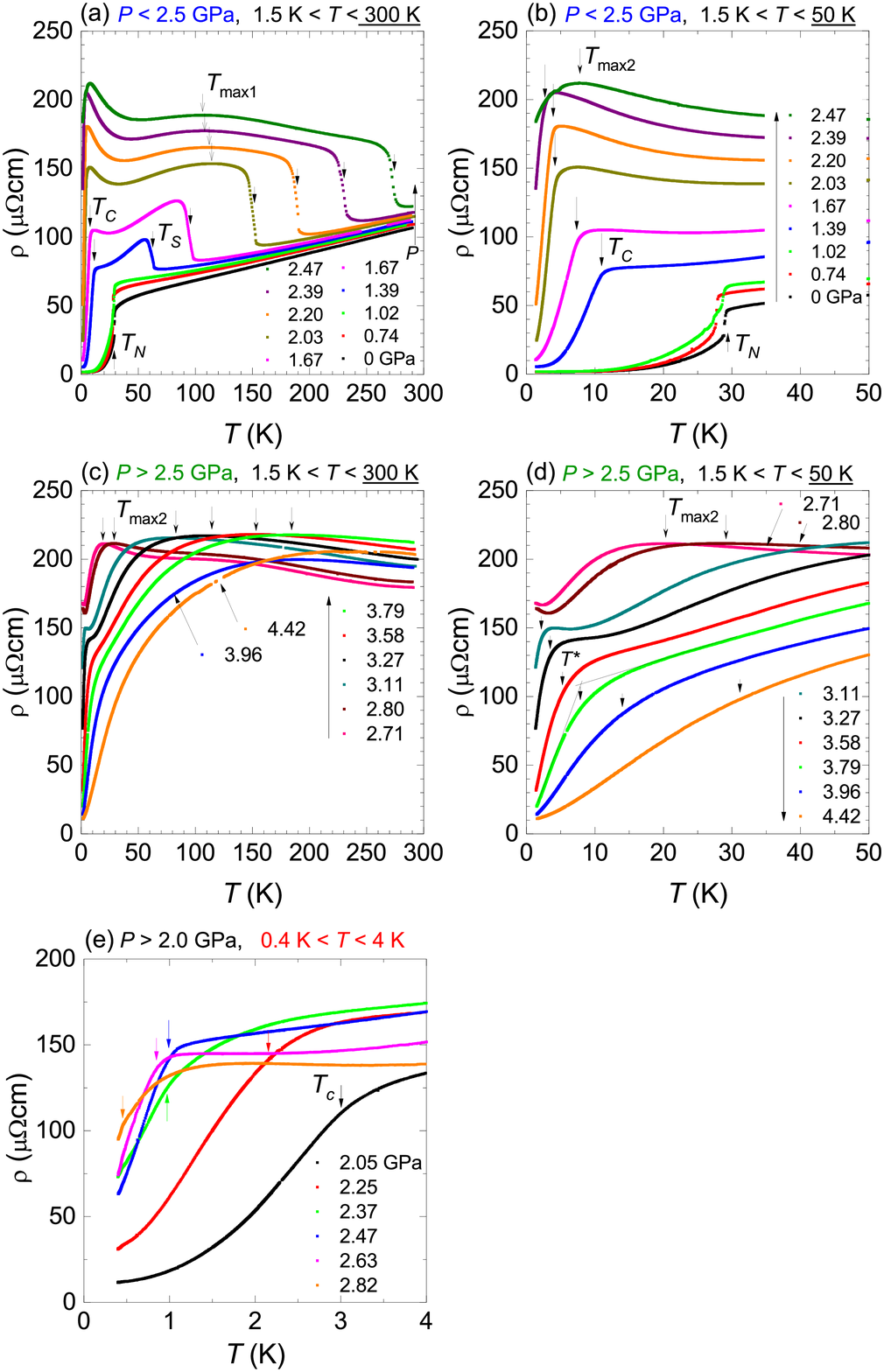}
\caption[]{(color online) Temperature dependence of the electrical resistivity for CeZn. (a), (b) $P<2.5$ GPa; (c), (d) $P>2.5$ GPa; and (e) the lower temperature part down to 0.4 K. The AFM transition changes into the FM transition in the tetragonal phase below $T_S$. The magnetically ordered state vanishes at $\sim3$ GPa. In the PM state, two local maxima denoted as $T_{\rm max1}$ and $T_{\rm max2}$ appear, and another characteristic temperature $T^*$ is seen above $\sim3$ GPa.}
\end{figure}

Figure 2(a-d) shows the temperature dependence of the electrical resistivity down to 1.5 K for several pressures: (a,b) $P<2.5$ GPa and (c,d) $P>2.5$ GPa. 
The low-temperature data down to 0.4 K are shown in Fig.~2(e).
The data of $P<2.5$ GPa in Fig.~2(a) and (b) are almost consistent with those of a previous study \cite{Kadomatsu}.
From ambient pressure to 1.02 GPa, the temperature dependence of the resistivity in the PM state shows monotonous variation, followed by the first-order AFM transition at $T_N\sim30$ K. 
The Kondo effect is unremarkable in this phase, and the ordered moment at ambient pressure has been reported to be 1.91 $\mu_B$/Ce \cite{Shigeoka}. 
Above 1.39 GPa, a first-order structural phase transition appears at $T_S$, as seen in the large jump of the resistivity.
In the tetragonal phase below $T_S$, the Kondo effect becomes prominent and the FM transition emerges.
First, we mention the pressure evolution of the resistivity in the tetragonal PM phase.
In the resistivity, two local maxima are seen in the range $2.03-2.47$ GPa.
One is $T_{\rm max1}\sim120$ K, which is insensitive to pressure.
This is likely related to the crystal electric field (CEF) splitting, which was 65 K between the ground-state quartet and the excited-state doublet in the cubic phase \cite{Pierre}.
The quartet is lifted into two doublets in the tetragonal phase, but the magnitude of the separation is unknown.
Another local maximum denoted as $T_{\rm max2}$ appears at low temperatures.
In Fig.~2(c), $T_{\rm max2}$ is approximately $10$ K at 2.71 GPa, and increases remarkably under further pressure and merges with $T_{\rm max1}$.
Another characteristic temperature denoted as $T^*$ appears, which is the shoulder in the resistivity at $\sim3$ GPa, as seen in Fig.~2(d). 
The $T^*$ is determined by a cross point of two extrapolation lines, as shown in the data at 3.79 GPa.
This is considered a signature of the coherent Kondo effect, which makes the ground state nonmagnetic and drastically increases under further pressure.
These characteristic temperatures are summarized in Fig.~1.
As mentioned above, $T_{\rm max1}$ is speculated to be related to the highest CEF level.
In the range $T >T_{\rm max1}$, the Kondo effect for all CEF states probably occurs.
In this context, $T_{\rm max2}$ may be related to the Kondo effect for the lower two doublets, that is, the quasiquartet.
Because it is lifted in the tetragonal phase, it is considered that $T^*$ appears as the Kondo effect for the ground-state doublet.

As seen in Fig.~2(b), the FM transition temperature $T_C$ is approximately 12 K at 1.39 GPa, which is consistent with that of previous studies \cite{Kadomatsu,Shigeoka}, and the neutron scattering measurement under pressure has confirmed that the FM transition in this pressure range is of the second order \cite{Shigeoka}.
The $T_C$ decreases gradually with increasing pressure, and it is suppressed below 1.5 K at 2.47 GPa.
Here the transition temperature was determined by the peak in $-d^2\rho/dT^2$.
The low-temperature data are shown in Fig.~2(e).
In this study, the transition temperature decreased continuously with decreasing pressure and reached the measurement limit at 2.82 GPa, which is consistent with that of a recent report \cite{XShen}. 
There is no clear indication that the transition changes into the first order.

\begin{figure}[tb]
\centering
\includegraphics[width=0.94\linewidth]{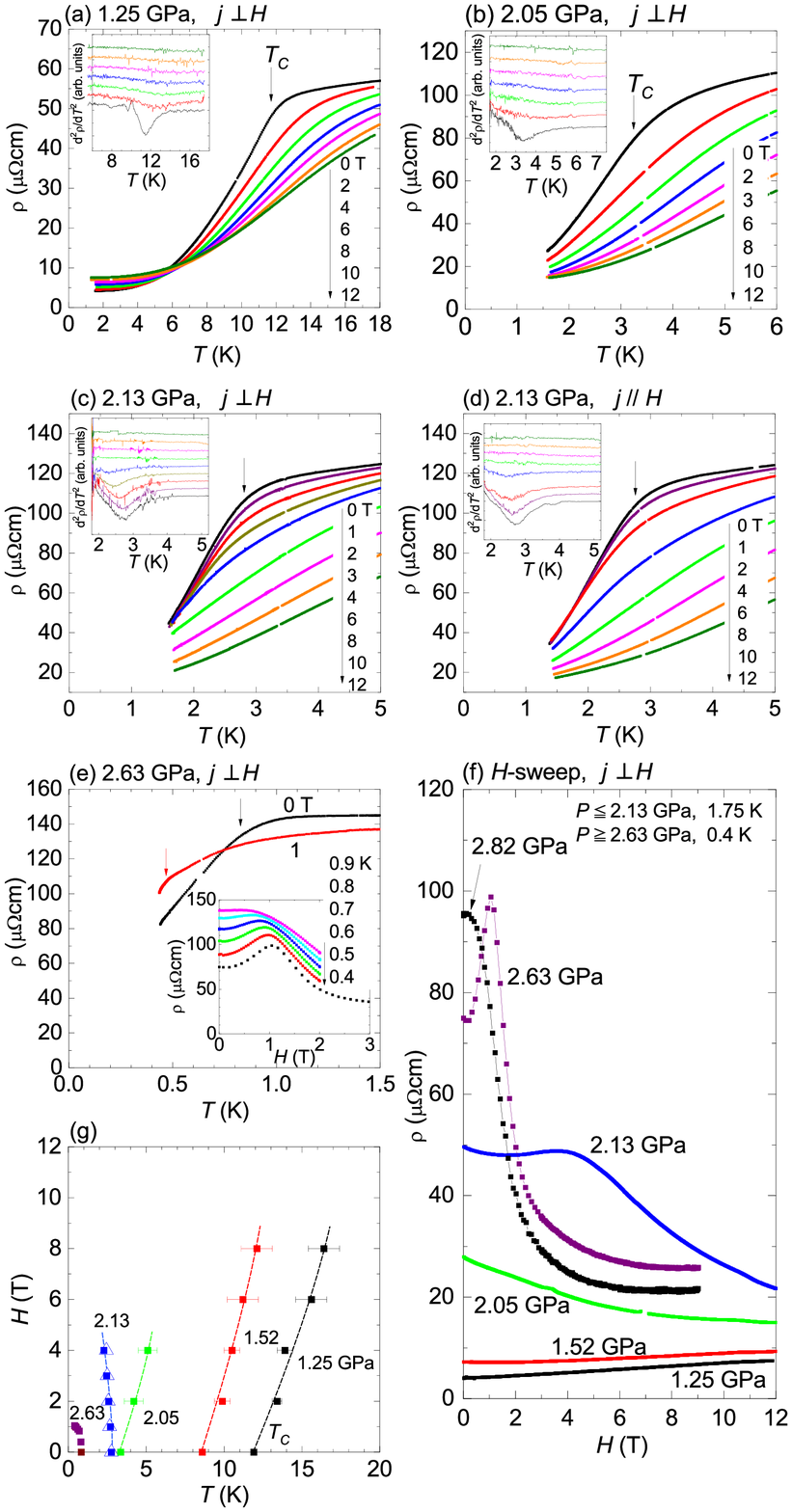}
\caption[]{(color online) (a-e) Temperature dependence of the resistivity under magnetic fields. At 1.25 and 2.05 GPa, the transition is changed into crossover as expected for an FM transition. Above 2.1 GPa, the transition temperatures decrease under magnetic fields, suggesting that the ordered phase is AFM-like. (c) and (d) show that the field-direction dependence is small at 2.13 GPa. Inset of (e) and (f): Corresponding metamagnetic-like transition appears in the field sweep of resistivity above 2.1 GPa, and they seem to reach 0 T at approximately 2.82 GPa. (g) The magnetic-field-temperature phase diagram of CeZn. The points for $j \parallel H$ at 2.13 GPa are shown by opened triangles.}
\end{figure}

To check whether or not the magnetically ordered state terminates at $P_c$ while maintaining the FM state, we investigated the field response of the ordered state.
Figures 3 (a-e) show the temperature variation of the resistivity under several magnetic fields at 1.25, 2.05, 2.13, and 2.63 GPa.
The $d^2\rho/dT^2$ is shown in the insets of Figs.~3(a-d).
At 1.25 GPa, where the FM state has been confirmed by the neutron scattering measurement \cite{Shigeoka}, the peak of $-d^2\rho/dT^2$ moves to higher temperatures under magnetic fields and the peak structure is broadened drastically.
This behavior is a typical one for FM systems, and the phase transition is suggested to be changed into a crossover under magnetic fields.
The behavior at 2.05 GPa is similar to that at 1.25 GPa.
However, at 2.13 GPa, the resistivity becomes insensitive to lower magnetic fields.
The peak of $-d^2\rho/dT^2$ is robust to the magnetic field and moves to lower temperatures, suggesting that the ordered state is no longer a simple FM state.
If the FM state exhibits strong anisotropy and the magnetic field is applied along the hard axis of the FM state, it is possible for this behavior to occur. 
To check this point, we changed the direction of the pressure cell against the magnetic fields at 2.13 GPa and confirmed the similar field response, as shown in Figs.~3(c) and (d).
This excludes a possibility that the FM state remains and the magnetic easy axis is changed above $\sim2.1$ GPa.
At 2.63 GPa, the transition temperature of $\sim0.82$ K at zero field clearly decreases to $\sim0.45$ K at 1 T.
Another important feature appears in the field sweep of the resistivity, which is shown in the inset of Fig.~3(e).
Obviously, the peak structures appear at approximately 1 T, indicative of the metamagnetic transition.

Figure 3(f) shows the field-sweep of the resistivity including the data under different pressures.
At lower pressures of 1.25, 1.52, and 2.05 GPa, there is no peak structure.
This is consistent with the interpretation that the ordered state in this pressure range is the FM state.
The peak structure starts to appear above 2.13 GPa and becomes obvious at 2.63 GPa.
The peak reaches 0 T at 2.82 GPa.
Figure 3(g) shows the magnetic field-temperature phase diagram in the tetragonal phase of CeZn.
The field response is drastically changed beyond $\sim2.1$ GPa.
Below 2.1 GPa, the temperature showing the anomaly increases, whereas it decreases above 2.1 GPa. 
At 2.63 GPa, the peaks of $-d^2\rho/dT^2$ and $\rho(H)$ are consistent with one another, and the magnetically ordered phase is closed under the magnetic field of $\sim1$ T.
Therefore, it is conjectured that the FM state is modified into the AFM-like state above $\sim2.1$ GPa.
Figure 1(b) shows the pressure-temperature phase diagram focusing on the magnetically ordered phase.
We distinguished the region, where the metamagnetic-like behavior is observed, as a different phase.
The phase boundary between the FM phase and the AFM-like phase is not clearly detected in this study.
We consider that the phase boundary between two ordered phases is of first order, because the field responses for two phases seem not to be connected smoothly under magnetic fields. 
If this is the case, the mixed state may be realized near 2.1 GPa, and the volume fraction of each phase is probably changed under magnetic fields.
The slope of the pressure dependence of the transition temperature changes at approximately 2.3 GPa.
This change indicates that the magnetostriction at the phase transition is different between the two regions, supporting the modification of the ordered state.

The switching of the ordered state occurs when the Curie temperature reaches approximately $3$ K.
The obtained phase diagram strongly impresses the difficulty of the realization of the FM QCP.
What prevents the FM QCP in CeZn is an interesting issue. 
In disordered systems, the spin-glass phase often appears just before the FM QCP is reached \cite{Brando}.
This should be excluded in CeZn, because CeZn is a stoichiometric and clean system, as demonstrated by the low residual resistivity.
A possible scenario is that an inhomogeneous magnetic phase is formed by quantum fluctuations \cite{Conduit,Efremov,Karahasanovic}.
The appearance of the new phase at low temperature implies a contribution of the quantum fluctuation for the formation of the phase.
The residual resistivity likely increases above 2.1 GPa, as shown in Fig.~2(e), suggesting that a complex magnetic structure is realized in the AFM-like phase.
Another scenario is the accidental change of the magnetic interaction (RKKY interaction) through the enhancement of the hybridization between the conduction electrons and $f$-electrons.
It is crucial to know the magnetic structure in the new phase to reveal the mechanism of switching of the phase.
It is also interesting to compare with the case in LaCrGe$_3$, where the magnetic correlations are insensitive irrespective of the change in the ground state \cite{Rana}.

In summary, we investigated the quantum phase transition in CeZn, in which the FM state appears in the pressure-induced tetragonal phase.
We observed the suppression of the magnetically-ordered phase under a pressure of $P_c \sim 3$ GPa.
However, the sudden change in the field response suggests that the FM state is changed into another AFM-like state before the QCP is reached.
The switching of the ordered state at low temperature suggests an interesting mechanism behind it.
CeZn is an example to understand how the FM QCP is avoided in the actual material.

\section*{Acknowledgements}

This work was supported by JSPS KAKENHI Grant Numbers JP15H05882, JP15H05885, JP18H04320, JP18H04321, and JP21K03446.

\end{document}